\documentclass[,final,numberedheadings]{aipproc}
\layoutstyle{6x9}

\begin{document}
\title{Hard Break-Up of Two-Nucleons and QCD Dynamics of NN Interaction}



\author{Misak Sargsian}{
address={Department of Physics, Florida International University, Miami, FL 33199}}

\begin{abstract}
We discus recent developments in theory of high energy two-body break-up 
of few-nucleon systems. The characteristics of these reactions 
are such that the hard two-body quasielastic subprocess can be clearly 
separated from the accompanying soft subprocesses. We discuss in details the 
hard rescattering model~(HRM) in which hard photodisintegration develops
in two stages.
At first, photon knocks-out an energetic quark which rescatters 
subsequently with a quark of the other nucleon.  
The latter provides a mechanism of sharing the initial high 
momentum  of the photon between two outgoing nucleons.  
This final state hard rescattering can be  expressed 
through the hard NN scattering amplitude. 
Within HRM  we discuss hard break-up reactions involving $D$ and $^3He$ targets 
and demonstrate how these reactions are sensitive to the dynamics of hard
$pn$ and $pp$ interaction.
Another development of HRM is the prediction of new helicity selection mechanism 
for hard two-body reactions, which was apparently  confirmed in the recent 
JLab experiment.
\end{abstract}

\maketitle

\section{Introduction to Nuclear QCD}
%There is  class of high-energy and momentum transfer (semi) exclusive 
%nuclear reactions which suits very well for studies of Nuclear QCD.  
There are several common features  in  theoretical approaches 
for studying QCD dynamics in hard nuclear reactions. These features 
are the following:
(a) one can clearly  identify a hard sub-process in the scattering mechanism; 
(b) which can be factorized from the soft nuclear part of the reaction;
(c) the subprocess is hard enough for  pQCD to be applicable; 
(d) soft part of the reaction could be expressed through  the measurable/extractable
quantities such as partonic distribution functions, hadron-hadron scattering 
amplitudes, form-factors and calculable nuclear wave functions.  
If all the above requirements are achieved such theoretical approaches 
may yield (sometimes) parameter free predictions (see e.g. \cite{gdpn}) and 
practically always will allow us to study the QCD aspects of strong interaction 
dynamics 
which otherwise can not be investigated without using nuclear targets.

The several of such reactions which satisfy  above criteria are, 
(i) semi-inclusive deep-inelastic  nuclear reactions aimed at studies of 
nuclear modification of partonic distribution functions~(PDFs) (EMC effects)
(see e.g. \cite{FS88,hnm,MeSaSi});
(ii) high momentum transfer elastic scattering off few-nucleon systems that can
be used to study the onset of quark degrees of freedom in strongly 
correlated few nucleon systems\cite{BCh}, 
(iii) hard electroproduction reaction  off nuclei
in which the final state interaction is 
used to probe the dominance of point like configurations in the hadronic 
wave function at large $Q^2$ (color transparency/coherence phenomena)
\cite{EFGMSS,FGMSS,gea,treview}; 
(iv) DIS nuclear scattering at $x_{Bj}>1$ as a framework for 
investigation of the mechanism of generation of 
super-fast quarks\cite{FS88,hnm} 
(v) as well as, hard break-up of two-nucleons in 
nuclei which can be used to probe the dynamics of strong interaction in 
two-nucleon system at intermediate to short distances\cite{gdpn,BCh,Carlson,GG02}.

\section{High energy break-up of two nucleons in nuclei}
Here  we focus on the last class (v) of the reactions, in 
which high energy photon produces two energetic nucleons which equally 
share the initial energy of the photon.  These reactions kinematically 
correspond to the break-up of 2N system at  $90^0$  angle in the 
$\gamma$-- $2N$ center of mass reference frame.

Due to completely symmetric kinematic configuration  the scattering process is 
dominated by photon  probing the structure and the dynamics of the exchanged 
particle in the NN system (Fig.1).  High energy  of photon in this case 
provides necessary resolution to probe  the QCD content of  NN interaction 
whether it proceeds through the $q\bar q$ exchange (Fig.1a), 
quark interchange (Fig.1b) or gluon exchanges (Fig.1c).

\begin{figure}
%\begin{center}
\centering\includegraphics[height=1.6in,width=3.5in]{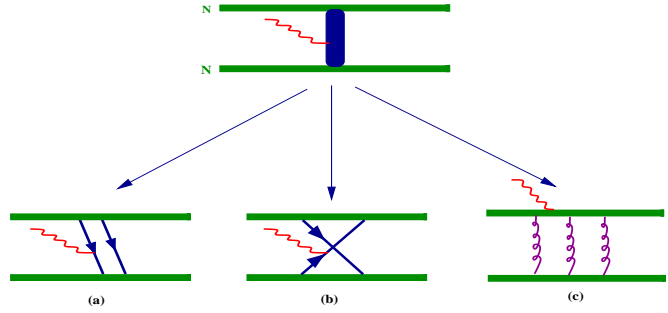}
%\psfig{file=figure1.eps,width=3.5in,height=1in}
%\end{center}
\caption{Possible QCD dynamics of NN interaction.}
\label{fig1}
\end{figure}

Effectiveness of these processes in probing QCD aspects of 
nuclear interaction  can be seen (for  $\gamma d\rightarrow pn$ reactions)
from  the following kinematical considerations\cite{BCh,Carlson,GG02}) in which:
\begin{equation}
s = (k_\gamma + p_d)^2 = 2M_dE_\gamma + M_d^2;  \ \ 
t = (k_\gamma - p_N)^2 = (cos(\theta_{cm})- 1){s-M_d^2\over 2}.
\label{kin}
\end{equation} 
Simple estimate shows that already at $E_\gamma > 2$~GeV the 
invariant momentum transfer $-t\mid_{90^0} > 4$~GeV$^2$  and 
invariant mass of the $NN$ system $M_{NN}=\sqrt{s} > 2$~GeV. 
These are conditions  for which one expects an onset of 
quark degrees of freedom  in the dynamics of strong interaction\cite{Feynman}.

One of the first theoretical predictions for high energy and large CM angle
break-up reactions based on quark-gluon content of interacting hadrons 
was the quark-counting rules\cite{BCh}. According to these rules it was 
predicted in Ref.\cite{BCh} that  the differential cross section of 
$\gamma d\rightarrow pn$ reaction in high momentum transfer and large CM angle  
regime should behave like ${d\sigma\over dt} \sim s^{-11}$. 
This prediction was experimentally 
confirmed already starting at $E_{\gamma}=1$~GeV for several set of 
experiments at SLAC\cite{NE8,NE17} and 
Jefferson Lab\cite{E89012,Schulte1,Schulte2,Mirazita}.

The quark counting  predictions are based on the hypothesis 
that the Fock states with  minimal number of partonic constituents 
dominate in two-body large angle hard collisions\cite{hex}. 
Although successful in describing energy dependences of the number of 
hard processes, this hypothesis does not allow to make calculation of 
the absolute values of cross sections. 
Especially for reactions involving baryons, calculations within 
perturbative QCD underestimate  the measured cross sections by the orders 
of magnitude (see e.g.\cite{Isgur_Smith}). 
This may be an indication that in the accessible range of energies 
the main part of the interaction is in the domain of nonperturbative
QCD\cite{Isgur_Smith,Rady}. However, the  problem  is that 
even if we fully realize the importance of  nonperturbative interactions 
the theoretical methods of calculations in the nonperturbative domain 
are very restricted.

\section{Hard Rescattering Mechanism of Two-Body Break-up Reactions}
%\section{Hard Rescattering Model} 

The underlying assumption in hard rescattering model~(HRM)\cite{gdpn} 
is that high energy photodisintegration of two-nucleon system proceeds 
through the two stages in which an absorption of  photon
by a quark of one  nucleon is  followed by  a high-momentum transfer
(hard) rescattering with a quark from the second nucleon. The 
latter rescattering produces a final two nucleon state with  large 
relative momenta. A typical diagram representing such a scenario is 
presented in Fig.2.

\begin{figure}[bh]
%\begin{center}
\centering\includegraphics[height=1.6in,width=3.3in]{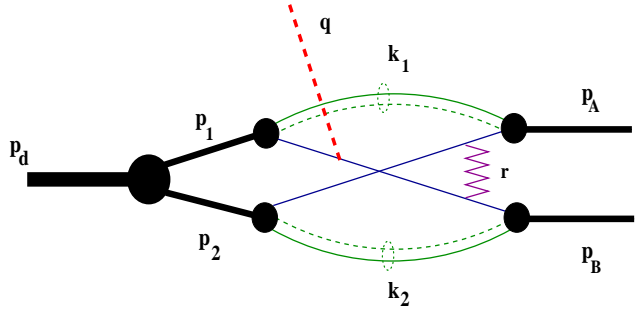}
%\psfig{file=figure2.eps,width=3.3in,height=0.9in}
%\end{center}
\caption{Typical diagram for hard rescattering mechanism.}
\label{fig2}
\end{figure}

Analyzing the type of diagrams as in Fig.2 allows us to do 
the following observations: 
\begin{itemize}
\item the dominant contribution comes from the soft vertices of 
$d\rightarrow NN$ transition,
while quark interchange rescattering  proceeds trough the hard 
gluon exchange,
\item the  $d\rightarrow NN$ transition can be evaluated through the 
conventional deuteron wave functions,
\item the structure of hard quark interchange interaction in the 
rescattering part of the reaction is similar to that of hard NN scattering,
\item as a result the sum of the multitude of diagrams with incalculable 
nonperturbative parts of the interaction can be  expressed through 
the experimentally measured amplitude of hard $NN$ scattering.
\end{itemize}
Based on these observations, calculation of the $\gamma + d\rightarrow pn$
amplitude yields\cite{gdpn,gdpnpol}
\begin{eqnarray}
& & \langle p_{\lambda_A},n_{\lambda_B}\mid A\mid \lambda_\gamma,\lambda_D\rangle = 
\sum\limits_{\lambda_2}  {f({\theta_{cm}})\over 3 \sqrt{2s'}}
\int \Psi^{\lambda_D,\lambda_\gamma,\lambda_2}(\alpha_c,p_\perp) {d^2p_\perp\over (2\pi)^2} \times \nonumber\\ 
& & \ \ \ 
\left(\langle p_{\lambda_A},{n_{\lambda_B}}|A_{pn}(s,t_n)|
p_{\lambda_\gamma},n_{\lambda_2}\rangle  - \right. %\nonumber \\
\left. \langle p_{\lambda_A},{n_{\lambda_B}}|A_{pn}(s,u_n)|
n_{\lambda_\gamma}p_{\lambda_2}\rangle \right),
\nonumber \\
\label{ampl}
\end{eqnarray}
where $A_{pn}$ is high momentum transfer elastic $pn$ scattering amplitude, 
$\mid N_{\lambda}\rangle$~($N=p,n$) represents the helicity wave function of 
nucleon and $\lambda_\gamma$ is the helicity of incoming photon.
Based on Eq.(\ref{ampl}) one obtains the following expression for the differential 
cross section of $\gamma + d\rightarrow pn$ reaction:
\begin{equation}
{d\sigma^{\gamma d \rightarrow pn}\over dt} = 
{8\alpha\over 9}\pi^4{1\over s^\prime}C({\tilde t\over s})
{d\sigma^{pn\rightarrow pn}\over dt}\left| \int 
\Psi_d^{NR}(p_z=0,p_t)\sqrt{m_n}{d^2p_t\over (2\pi)^2}\right|^2,
\label{crs_gdpn}
\end{equation}
where $s^\prime = s - 4m_N^2$ and $\tilde t = (p_n-m_n)^2$. The 
interesting properly of the function $C$ is that 
$C(\theta_{cm}=90^0) \approx 1$. Therefore for $90^0$~CM scattering    
HRM prediction is parameter free. Since $pn$ cross section at high 
momentum transfer behaves like $s^{-10}$, the additional ${1\over s^\prime }$ factor
in Eq.(\ref{crs_gdpn})  yields
the same $s^{-11}$ dependence for $\gamma + d\rightarrow p+n$ differential cross 
section as it follows from quark counting rule. However above derivation 
does not require an onset of pQCD regime. Also, due to 
angular dependence of the $pn$ amplitude, HRM predicts  
an angular distribution being not symmetric around $90^0$~CM. 
These predictions agree reasonably well  with the experimental 
data\cite{E89012,Schulte2}(see e.g. Fig.3).

\begin{figure}[bh]
%\begin{center}
\centering\includegraphics[height=3in,width=2.8in]{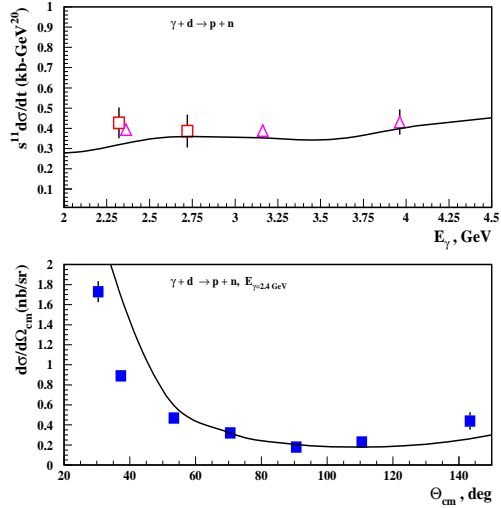}
%\psfig{file=figure3.eps,width=2.8in,height=1.8in}
%\end{center}
\caption{Energy dependence of the scaled cross section at $90^0$ CM scattering (top) 
and angular dependence of the cross section at $E_\gamma = 2.4$~GeV  
(bottom).}
\label{fig3}
\end{figure}

\section{Hard break-up of two protons from ${\bf ^3He}$}
With all its success and accuracies yet to be improved  HRM is only one of 
the approaches in describing hard photodisintegration reactions. Other 
models such as reduced nuclear amplitude~(RNA) formalism\cite{RNA} and 
quark-gluon string~(QGS) model\cite{QGS} describe many features of hard 
photodisintegration reaction, with QGS being rather successful in describing 
lower energy data.  However RNA and QGS require an absolute normalization.

Recently it was suggested\cite{ghppn0,ghppn} that the break-up of pp pair from 
$^3He$ will further advance our understanding of the dynamics of 
hard photodisintegration of two-nucleon systems and 
allow further discrimination between above mentioned models~(see also Ref.\cite{EIPtalk}).

Within HRM the typical diagram describing two-proton break-up is shown in Fig.4.

\begin{figure}
%\begin{center}
\centering\includegraphics[height=1.6in,width=3.2in]{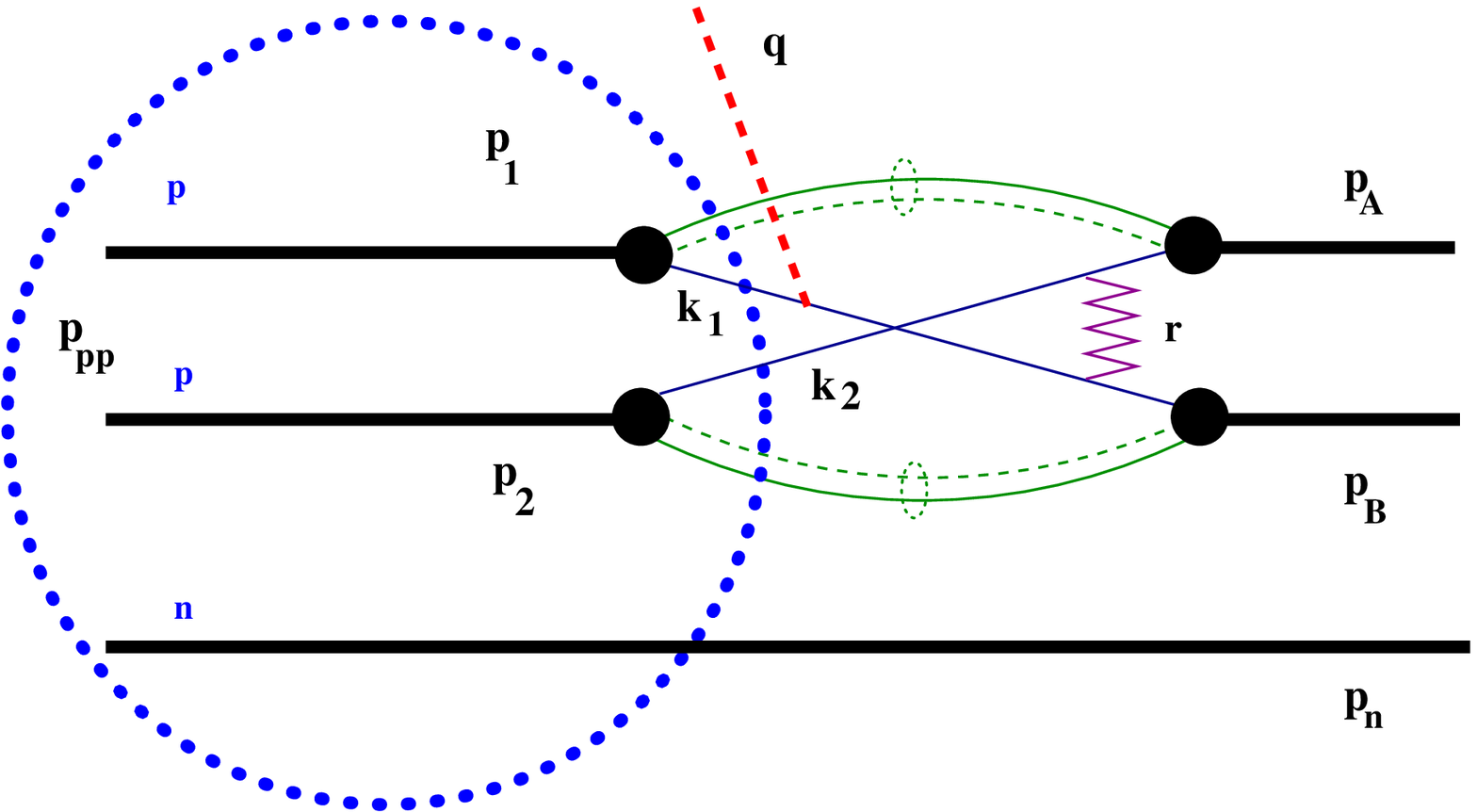}
%\psfig{file=figure4.eps,width=3.2in,height=1.2in}
%\end{center}
\caption{Typical diagram for hard break-up of pp pair from $^3He$.}
\label{fig4}
\end{figure}
HRM calculation similar to that of deuteron  break-up reaction yields:
\begin{equation}
{d\sigma\over dt d^3p_n} = \left({14\over 15}\right)^2 
{16\pi^4 \alpha \over S-M_{^3He}^2} ({2c^2\over 1+2c^2})
{d\sigma^{pp}\over dt} (s_{pp},t_n){S_{34}\over E_n},
\label{pp}
\end{equation}
where 
\begin{equation}
S_{34} = \sum\limits_{\lambda_1=-\lambda_2,\lambda_3=-{1\over 2}}^{1\over 2} 
\left | \int \psi^{1\over 2}_{^3He}(\lambda_1,\lambda_2,\lambda_3)m {d^2 p_{2\perp}\over (2\pi)^2}\right |^2,
\end{equation}
and $c={|\phi_{3,4}|\over |\phi_1|}$ with $\phi_i$ being $pp$ 
helicity amplitudes. Since the cross section of hard $pp$ scattering enters in 
Eq.(\ref{pp}) one of the interesting predictions of HRM is the possibility of  
observation of energy oscillations at $90^0$ CM scattering similar to one observed 
in elastic $pp$ scattering. Another interesting feature of two-proton break-up 
reactions
is the fact that at lower energies this reaction is three-step\cite{Laget} rather than 
two-step process. HRM in fact predicts that with an increase of photon energy due 
to the onset of quark-interchange (rather than meson-exchange) mechanism of 
NN interaction the   
two-body processes will dominate the cross section. The pioneering 
experiment of high energy $pp$ break-up reaction\cite{roneip} 
was recently performed at Jefferson Lab which 
may shed new light on  many issues of  hard rescattering processes. 

\section{Helicity Transfer Mechanism in Two-Body Break-up Reactions}
In addition to the cross section measurements, polarization observables may provide 
a new insight into the dynamics of hard photodisintegration. Original 
motivation for polarization measurements in high energy photodisintegration reaction 
was the expectation that the onset of the pQCD regime in the reaction dynamics 
will be accompanied by an observation of the helicity conservation in polarized 
reactions. Both energy and angular distributions of several polarization observables 
have been measured at JLab\cite{gdpnpolexp1,gdpnpolexp2}. Although the energy range 
covered was rather restricted, it provided an interesting insight into the  
structure of HRM.

One of the unique features of HRM  is that the struck quark carries 
the helicity of incoming photon.  As a result one of the final nucleons will carry 
the bulk of the polarization of incident photon (see e.g. Eq.(\ref{ampl})).  
Thus in HRM photon plays as a helicity selector for the final nucleons.  
This yields  a prediction for large asymmetry\cite{roneip}~($C_{z'}$) 
for the longitudinal polarization of outgoing nucleons.  In Ref.\cite{gdpnpol} 
we predicted a sizable asymmetry for $C_{z'}$ even though the 
existing data\cite{gdpnpolexp1} (with rather large errors)  at that  
time were indicating on vanishing values of $C_{z'}$.  

\begin{figure}[bh]
%\begin{center}
\centering\includegraphics[height=3in,width=3.4in]{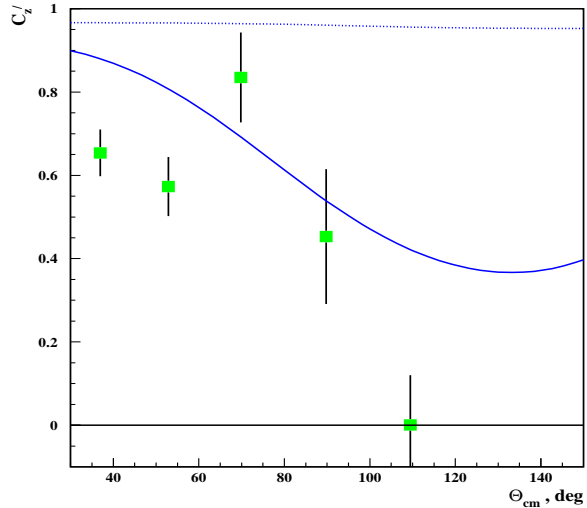}
%\psfig{file=figure5.eps,width=3.4in,height=1.6in}
%\end{center}
\caption{Angular dependence of $C_{z'}$ for $E_\gamma = 1.9$~GeV.}
\label{fig5}
\end{figure}
However, recent data\cite{gdpnpolexp2}, appears to confirm HRM prediction 
for  large values of $C_{z'}$ (see Fig.5). It will be interesting also 
to check the other HRM prediction that $C_{z'}$ will continue to approach to unity 
with an increase of  photon energy at $90^0$ CM scattering.

It is very interesting that above described helicity selection mechanism of HRM 
predicts ({\em an opposite}) vanishing value of $C_{z'}$ for two-body break up of proton 
pair from $^3He$.  This follows from the fact that the dominant part of the 
amplitude which represents  two final state nucleons polarized in same direction 
are proportional to the nuclear ground state wave function with two initial  nucleons 
having same helicities.  Due to Pauli principle this part of the amplitude is strongly 
suppressed for the proton-pair in $^3He$ target. No such suppression exists for 
$pn$ break up reactions.

\section{Summary and Outlook}
There is an accumulating evidence that 
hard rescattering mechanism explains the underlying dynamics of high energy 
and large CM angle break-up of a nucleon pair from $D$ and $^3He$ targets.
One of the important features of HRM is that its prediction of unpolarized 
cross section at $90^0$ center of mass photodisintegration of deuteron is 
parameter free and no further adjustments are required.

HRM predicts that energy dependence of  two-proton break-up reaction should 
resemble that of hard elastic pp cross section, which could mean an observation of 
oscillation for $s$-weighted $\gamma + ^3He\rightarrow (pp) + n$ cross section similar to 
the one observed in hard elastic $pp$ scattering.

Another feature of HRM, observed recently, is the prediction of large longitudinal 
asymmetries in $\gamma d\rightarrow pn$ reactions due to the helicity transfer 
mechanism characteristic to hard  rescattering model.

If HRM will prove to be a true mechanism of hard photodisintegration 
reaction involving two nucleons, it will advance also our understanding 
of the dynamics of NN interaction at short distances.

A new venue for advancing our understanding of the dynamics of hard break-up 
reactions could be an extension of these studies to the kinematics in 
which two excited baryonic states (like $\Delta$-isobars) are produced 
at large center of mass angles of $\gamma$ -- $NN$ system.

%\begin{verbatim}

%\end{verbatim}

\end{document}